\documentclass[aps,prl,floats,floatsfix,twocolumn,superscriptaddress,showpacs]{revtex4-1}

\usepackage{dcolumn} 
\usepackage{bm} 
\usepackage{amsthm}
\usepackage{latexsym}
\usepackage{float}
\usepackage{amsfonts}
\usepackage{amsmath}
\usepackage{amssymb}
\usepackage{color,graphicx}

\begin{document}
\title{Exciton-Trion Coherent Interactions in a CdTe/CdMgTe Quantum Well}

\author{G.~Moody}
\affiliation{JILA, University of Colorado \& National Institute of Standards and Technology, Boulder CO 80309-0440}
\affiliation{Department of Physics, University of Colorado, Boulder CO 80309-0390}
\author{I.~A.~Akimov}
\affiliation{Experimentelle Physik 2, Technische Universit$\ddot{\textit{a}}$t Dortmund, D-44221 Dortmund, Germany}
\affiliation{A. F. Ioffe Physical-Technical Institute, Russian Academy of Sciences, 194021 St. Petersburg, Russia}
\author{H.~Li}
\affiliation{JILA, University of Colorado \& National Institute of Standards and Technology, Boulder CO 80309-0440}
\author{R.~Singh}
\affiliation{JILA, University of Colorado \& National Institute of Standards and Technology, Boulder CO 80309-0440}
\affiliation{Department of Physics, University of Colorado, Boulder CO 80309-0390}
\author{D.~R.~Yakovlev}
\affiliation{Experimentelle Physik 2, Technische Universit$\ddot{\textit{a}}$t Dortmund, D-44221 Dortmund, Germany}
\affiliation{A. F. Ioffe Physical-Technical Institute, Russian Academy of Sciences, 194021 St. Petersburg, Russia}
\author{G.~Karczewski}
\affiliation{Institute of Physics, Polish Academy of Sciences, PL-02668 Warsaw, Poland}
\author{M.~Wiater}
\affiliation{Institute of Physics, Polish Academy of Sciences, PL-02668 Warsaw, Poland}
\author{T.~Wojtowicz}
\affiliation{Institute of Physics, Polish Academy of Sciences, PL-02668 Warsaw, Poland}
\author{M.~Bayer}
\affiliation{Experimentelle Physik 2, Technische Universit$\ddot{\textit{a}}$t Dortmund, D-44221 Dortmund, Germany}
\author{S.~T.~Cundiff}
\email{cundiff@jila.colorado.edu}
\affiliation{JILA, University of Colorado \& National Institute of Standards and Technology, Boulder CO 80309-0440}
\affiliation{Department of Physics, University of Colorado, Boulder CO 80309-0390}

\begin{abstract}
We present a collection of zero-, one- and two-quantum two-dimensional coherent spectra of excitons and trions in a CdTe/(Cd,Mg)Te quantum well.  The set of spectra provides a unique and comprehensive picture of the exciton and trion nonlinear optical response.  Exciton-exciton and exciton-trion coherent coupling is manifest as distinct peaks in the spectra, whereas signatures of trion-trion interactions are absent.  Excellent agreement using density matrix calculations is obtained, which highlights the essential role of many-body effects on coherent interactions in the quantum well.
\end{abstract}

\date{\today}
\maketitle

The coherent optical response of collective resonances is often influenced by interactions between the individual oscillators.  Resonant interactions appear in a wide range of physical and chemical systems, including optical Feshbach resonances in ultracold atomic gases \cite{Blatt2011}, chemical reactions of ultracold polar molecules \cite{Miranda2011}, dipole-dipole interactions in dilute atomic vapors \cite{Dai2012}, nuclear spin-spin coupling \cite{Ernst1987} and excitonic effects in light-harvesting reaction centers \cite{Engel2007,Collini2010}.  In the solid-state, spin phenomena in semiconductor nanostructures have garnered considerable interest in recent years for potential applications in spintronic devices and quantum information processing \cite{Awschalom2013}.  Among the various ensemble spin systems, a two-dimensional electron gas (2DEG) in a modulation-doped quantum well (MDQW) is particularly interesting because a dense spin ensemble exhibiting little to no inhomogeneity is readily grown using epitaxial methods.  At low temperature, the band-edge optical properties of a MDQW are dominated by Coulomb-bound electron-hole pairs (excitons) and charged excitons (trions), analogous to $H$ and $H^-$ or $H^+_2$, respectively.

Exciton and trion resonances have been exploited for a variety of spin phenomena, including long-lived electron and hole spin oscillations \cite{Chen2007,Zhukov2009}, coherent spin rotations about the Bloch sphere \cite{Phelps2009}, optical quantum memories \cite{Langer2012} and electromagnetically-induced transparency (EIT) \cite{Wang2012}.  These processes rely on manipulation of the 2DEG through coherent light-matter interactions of the exciton and trion transitions, whose optical properties are strongly influenced by many-body effects (MBEs) inherent to semiconductors.  For example, the fidelity of collective electron spin rotations is hindered by exciton-trion interactions, and excitation of the exciton can reduce the polarized 2DEG spin coherence time by an order of magnitude \cite{Phelps2009}.  Additionally, Coulomb interactions can limit the level of achievable transparency in EIT experiments using an electron-trion $\Lambda$-system to less than a few percent \cite{Wang2012}.  These examples illustrate the importance of MBEs on the coherent optical response of MDQWs, which have not been adequately characterized.  Moreover, coherent coupling between excitons and trions is interesting, because  the complexes need to be in close proximity for it to become significant. Typically, trions are localized at cryogenic temperatures, while excitons can be spatially more extended \cite{Astakhov2002}.  The effective spatial overlap of the wavefunctions of these states determines their coupling strength.  Thus, establishing the influence of MBEs on the nonlinear optical properties of MDQWs is imperative for facilitating development of spin-based devices as well as for enhancing our understanding of fundamental interactions between neutral and charged particles residing in a plasma.

Linear spectroscopies provide some insight in this regard, revealing three-particle interactions that govern trion formation dynamics \cite{Combescot2005,Oberli2009} and oscillator-strength-stealing phenomena \cite{Brinkmann1999,Plochocka2004}.  Nonlinear spectroscopies such as transient absorption and four-wave mixing (FWM) techniques are sensitive to changes in the optical response that occur when particles interact through Coulomb forces \cite{Meier2007} or local fields \cite{Leo1990}.  These techniques have been used to probe for signatures of exciton-trion correlations stemming from phase-space filling and fermionic exchange \cite{Wagner1999,Oberli2004}; however, MBEs such as excitation-induced dephasing (EID) \cite{Wang1993} and excitation-induced energy shift (EIS) \cite{Shacklette2002} cannot be reliably distinguished using one-dimensional techniques since the numerous quantum pathways contributing to the nonlinear optical response are not sufficiently separated \cite{Mukamel2000,Shacklette2002}.

In this Letter, we use optical two-dimensional coherent spectroscopy (2DCS) \cite{Cundiff2012} -- an enhanced version of three-pulse transient FWM -- to overcome these limitations, thus providing unique insight into the MBEs arising from exciton-exciton and exciton-trion interactions.  We present a set of 2D spectra measured from a nominally undoped CdTe/(Cd,Mg)Te QW, which has not been previously studied using 2DCS methods.  Specific quantum pathways are isolated by measuring 2D spectra generated using different pulse time orderings, enabling differentiation between the many-body contributions to the coherent nonlinear optical response \cite{Li2006}.  Each type of 2D spectrum better separates the quantum pathways associated with interactions in the system compared to its one-dimensional counterparts; however we demonstrate that only when the collection of different types of 2D spectra are considered can a comprehensive picture of the nonlinear optical response be established.  Excellent agreement between density matrix calculations and the measurements reveals the essential role of MBEs in coherent excitonic interactions in the QW.  Distinct peaks in the one-quantum spectrum -- which correlates the excitation and emission energies of the system -- indicate the presence of exciton-trion interactions that arise predominantly through an EIS of their correlated state.  A two-quantum 2D spectrum -- correlating the non-radiative-double-quantum and emission energies -- has been calculated and proven to be particularly sensitive to MBEs \cite{Dai2012,Yang2008a,Stone2009,Karaiskaj2009} and reveals that interactions between trions are negligible due to spatial separation, whereas exciton-exciton and exciton-trion interactions are coherent.  The measurements and analysis presented here are essential for understanding interactions between neutral and charged particles in the solid state and might facilitate improvement or optimization of devices relying on coherent interactions between an ensemble of oscillators.

\begin{figure}[h]
\centering
\includegraphics[width=0.9\columnwidth]{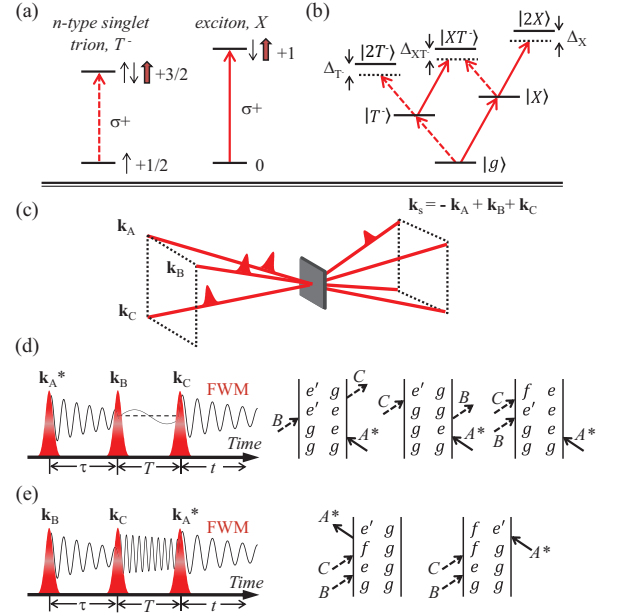}
\caption{(Color online) (a) The exciton and negative trion optical transitions accessible using circularly polarized light ($\sigma+$).  The exciton consists of a spin $-1/2$ electron (thin arrow) and spin $+3/2$ heavy hole (thick arrow), whereas the trion consists of two electrons in a spin $0$ singlet state correlated with a spin $+3/2$ heavy hole. (b) The exciton and trion nonlinear response is modeled using a six-level energy scheme.  The energy diagram consists of a ground state ($|g\rangle$), singly-excited exciton ($|X\rangle$) and trion ($|T^{-}\rangle$) states, doubly-excited exciton ($|2X\rangle$) and trion ($|2T^{-}\rangle$) states shifted from the singly-excited transition energies by $\Delta_X$ and $\Delta_{T^{-}}$, respectively, and a doubly-excited mixed exciton-trion state ($|XT^{-}\rangle$) shifted from the exciton + trion energy by $\Delta_{XT^{-}}$. (c) Geometry of the incident beams and the FWM signal. The pulse time ordering and generalized double-sided Feynman diagrams are shown for the (d) rephasing zero- and one-quantum sequence and the (e) two-quantum sequence.  The states $|e\rangle$ and $|e^{'}\rangle$ represent either $|X\rangle$ or $|T^{-}\rangle$ and $|f\rangle$ represents either $|2T^{-}\rangle$, $|2X\rangle$ or $|XT^{-}\rangle$.}
\label{fig1}
\end{figure}

The sample consists of a single 20 nm wide CdTe/CdMgTe QW grown by molecular beam epitaxy on a (100)-oriented GaAs substrate.  The QW is separated from the substrate by a CdTe/CdMgTe super-lattice grown on top of a thick CdMgTe buffer layer and is separated from the surface by a 100 nm CdMgTe barrier.  The sample is nominally undoped but due to residual impurities and charge redistribution to surface states, the QW at low temperature contains a dilute 2DEG (verified through magneto-photoluminescence spectra, data not shown).  The sample is mounted on a sapphire disk and the substrate is chemically-removed for transmission experiments.  Optical excitation generates excitons and trions with total angular momentum projections along the growth direction of $J_{X} = \pm1$ and $J_{T^{-}} = \pm3/2$, respectively.  The optically-active transitions accessible using circularly polarized light ($\sigma+$) are shown in Fig. \ref{fig1}(a) for the exciton, comprised of a spin $J_{e} = -1/2$ electron (thin arrow) and spin $J_{h} = +3/2$ heavy hole (thick arrow), and for the negative trion, which consists of two opposite spin electrons in a singlet state correlated with a spin $J_{h} = +3/2$ heavy hole.

Optical 2DCS experiments are performed using four phase-stabilized pulses propagating in the box geometry \cite{Bristow2009}.  The pulses, obtained from a mode-locked laser operating at a 76 MHz repetition rate, have a $\sim$ 150 fs duration and are all co-circularly polarized.  Three of the pulses $A$, $B$ and $C$ with wavevectors $\textbf{k}_{A}$, $\textbf{k}_{B}$ and $\textbf{k}_{C}$, respectively, are focused to a single $\sim$ 50 $\mu$m spot on the sample, which is kept at a lattice temperature of 6 K in a cold finger helium cryostat.  The exciton and trion excitation densities are kept below $\sim 5 \times 10^{9}$ cm$^{-2}$ so that the coherent nonlinear optical response is in the $\chi^{(3)}$ regime.  The pulses interact nonlinearly with the sample to generate a FWM signal that is detected along the phase-matched direction $\textbf{k}_{s}=-\textbf{k}_{A} + \textbf{k}_{B} + \textbf{k}_{C}$, which necessarily requires that pulse $A$ acts as a conjugate pulse irrespective of pulse time ordering, as shown in the schematic diagram in Fig. \ref{fig1}(c).  The signal is heterodyned with a phase-stabilized reference pulse and their interference is spectrally-resolved with $\sim 20$ $\mu$eV resolution.  For a rephasing experiment, interferograms are measured while the delay $\tau$ between the first two pulses incident on the sample, $A$ and $B$, is scanned with interferometric precision, as shown in the timing sequence in Fig. \ref{fig1}(d).  The spectrally-resolved FWM signal is Fourier transformed with respect to $\tau$ to generate a rephasing one-quantum spectrum that correlates the excitation and emission energies for a fixed delay $T=200$ fs.  Alternatively, the delay $T$ between pulses $B$ and $C$ can be scanned while $\tau$ is held fixed at 200 fs, and the signal can be Fourier-transformed with respect to $T$ to generate a rephasing zero-quantum spectrum, revealing population decay dynamics and non-radiative coherent superpositions between states \cite{Yang2008b,Dai2010,Moody2013}.  Moreover, the pulse time ordering can be adjusted so that the conjugated pulse $A$ is incident on the sample last, as depicted in the timing sequence in Fig. \ref{fig1}(e).  Analogous to negative delay two-pulse FWM experiments, the FWM signal generated using this timing sequence appears only if MBEs are present in the sample \cite{Yang2008a,Stone2009,Karaiskaj2009}.  The delay $\tau$ between pulses $B$ and $C$ is held fixed at 200 fs while the delay $T$ is scanned, and the signal is Fourier-transformed with respect to $T$ to generate a two-quantum spectrum that correlates the two-quantum excitation energies with the one-quantum emission energies.

\begin{figure}[h]
\centering
\includegraphics[width=0.95\columnwidth]{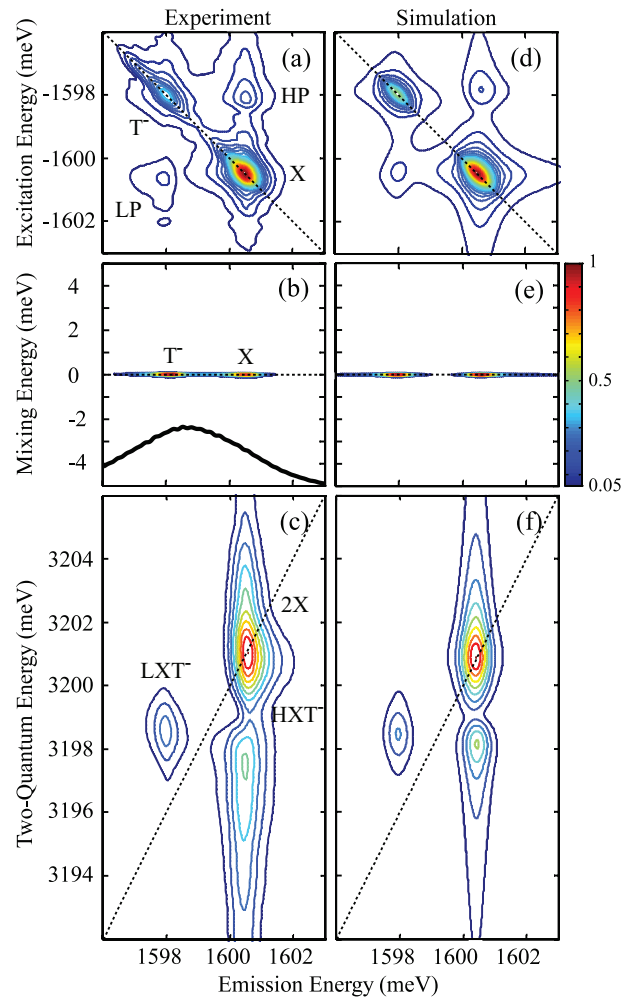}
\caption{(Color online) Normalized experimental rephasing (a) one-quantum and (b) zero-quantum spectra of the exciton ($X$), trion ($T^-$) and their interaction ($LP$ and $HP$).  Two-quantum coherences of the exciton ($2X$) and the mixed exciton-trion state ($LXT^{-}$ and $HXT^{-}$) appear in the experimental two-quantum spectrum shown in (c).  The excitation laser spectrum for all experiments is shown in the inset to (b).  Panels (d)-(f) are the corresponding density matrix simulations.  The color bar indicates the normalized amplitude of each spectrum.}
\label{fig2}
\end{figure}

Figure \ref{fig2}(a) shows the absolute value of the rephasing one-quantum spectrum.  We note that the vertical axis is plotted as negative excitation photon energy since the coherences created by the conjugated pulse oscillate at negative frequencies during $\tau$ with respect to the coherences during $t$.  The spectrum features two peaks on the diagonal corresponding to excitation and emission at the exciton ($X$) and trion ($T^{-}$) transitions.  The cross-diagonal and diagonal slices of each peak are simultaneously fit to analytical functions to determine the homogeneous and inhomogeneous line widths \cite{Siemens2010}.  The $X$ and $T^{-}$ homogeneous line widths are $\approx 0.15$ meV and $\approx 0.1$ meV, respectively, and the inhomogeneous line width of both peaks is $\approx 0.5$ meV.  The trion resonance is red-shifted from the exciton by a 2.7 meV binding energy.  The appearance of two off-diagonal peaks ($LP$ and $HP$) is a signature of quantum mechanical coupling between the exciton and trion, since they indicate excitation at one transition energy and emission at the other.

Cross peaks in rephasing spectra corresponding to coherent coupling are often accompanied by peaks in the rephasing zero-quantum spectrum associated with non-radiative Raman-like coherences between the transitions \cite{Yang2008b}; thus measuring this spectrum is a natural way to probe for coherent interactions.  A rephasing zero-quantum spectrum is shown in Fig. \ref{fig2}(b), which features two peaks at zero mixing energy and at the trion ($T^{-}$) and exciton ($X$) emission energies, corresponding to the system being in a ground or excited state population during the delay $T$.  Peaks at a mixing energy equal to $\pm2.7$ meV -- a clear signature of coherent coupling between resonances -- are absent, which would suggest that the exciton-trion interactions are incoherent.  On the other hand, the presence of cross-peaks ($LXT^{-}$ and $HXT^{-}$) associated with a collective exciton-trion two-quantum coherence in the two-quantum spectrum in Fig. \ref{fig2}(c) necessarily implies that the exciton and trion interact in a coherent manner.  The peak ($2X$) on the diagonal arises from coherent interactions between two excitons in the QW.  The absence of a two-trion peak, which would appear on the diagonal line at a two-quantum energy equal to twice the trion transition energy, indicates that coherent coupling between trions is absent due to spatial separation.

The collection of the different types of 2D spectra provides a unique perspective into the coherent nonlinear optical response of excitons, trions and a 2DEG residing in a QW.  To better understand the effects of exciton-trion interactions, we simulate the spectra by analytically solving a perturbative expansion of the density matrix for a six-level system, shown in Fig. \ref{fig1}(b).  The energy scheme consists of a ground state ($|g\rangle$), singly-excited exciton ($|X\rangle$) and trion ($|T^{-}\rangle$) states, doubly-excited states representing exciton-exciton ($|2X\rangle$) and trion-trion ($|2T^{-}\rangle$) correlations, and a doubly-excited mixed exciton-trion state ($|XT^{-}\rangle$).  In the absence of MBEs, such a level diagram is equivalent to four independent two-level systems through a Hilbert space transformation \cite{SM}.  MBEs can be introduced by breaking the equivalence of the ground state $\leftrightarrow$ singly-excited state transitions with respect to the singly-excited state $\leftrightarrow$ doubly-excited state transitions.  The effects of EID and EIS are modeled by altering the dephasing rate and transition energy, respectively, of the upper transitions compared to the lower transitions, as suggested in Ref. \cite{Bott1993}.

The quantum pathways that contribute to the nonlinear optical response are characterized by the double-sided Feynman diagrams in Figs. \ref{fig1}(d) and \ref{fig1}(e), which are written in a generalized form for which the states labeled with $|e\rangle$ and $|e'\rangle$ can be replaced with $|X\rangle$ or $|T^{-}\rangle$, and the state labeled by $|f\rangle$ with $|2X\rangle$, $|2T^{-}\rangle$ or $|XT^{-}\rangle$.  Expanding the diagrams in Fig. \ref{fig1}(d) results in 14 quantum pathways that contribute to the rephasing zero- and one-quantum spectra.  Similarly for the two-quantum spectrum, the diagrams in Fig. \ref{fig1}(e) can be expanded into 12 pathways.  Perturbation calculations are performed using Dirac delta function pulses in time.  Inhomogeneity that allows for uncorrelated broadening between transitions is included by integrating the third-order polarization over a Gaussian distribution of transition frequencies \cite{Cundiff1994}.  The homogeneous and inhomogeneous line widths are adjusted to match the measurements.  The coefficient characterizing the level of correlation between transition energy fluctuations, $R$, is set equal to zero for all pathways that involve both the exciton and trion transitions \cite{Moody2013}; otherwise it is set equal to unity.  The model cannot account for the influence of finite bandwidth of the excitation pulses on the peak amplitudes in the measurements.  Nonetheless, the amplitudes are matched by setting the optical dipole moment of the trion transition equal to $80\%$ of the exciton transition for all simulated spectra.

Simulated spectra are shown in the right column of Fig. \ref{fig2}.  The measurements are only reproduced for one specific set of parameters for all spectra.  We would like to stress this point: the complete collection of 2D measurements is necessary to provide enough constraints to identify the type of couplings in the system.  Without sufficient separation of the quantum pathways, either by analyzing only a subset of the spectra or probing the sample using one-dimensional methods, a comprehensive picture of the coherent nonlinear optical response cannot be established.  The simulations demonstrate that the inclusion of MBEs is essential to model the experimental data.  Without them, the off-diagonal cross peaks ($LP$ and $HP$) in Figs. \ref{fig2}(a) and \ref{fig2}(d) would be absent and the two-quantum signal would be zero.

Comparison of the simulations to the experiment reveals that the cross peaks in both the one- and two-quantum spectra originate from an EIS of the mixed $|XT^{-}\rangle$ state equal to $\Delta_{XT^{-}} \approx \pm 50$ $\mu$eV.  Similarly, the $2X$ peak in Fig. \ref{fig2}(c) stems from an EIS equal to $\Delta_X \approx \pm 0.12$ meV.  The two-quantum coherence line widths are reproduced by setting the $|g\rangle \rightarrow |2X\rangle$ and $|g\rangle \rightarrow |XT^{-}\rangle$ dephasing rates equal to $\approx 0.3$ meV and $\approx 0.2$ meV, respectively, indicating that the correlated states dephase in a picosecond timescale.  The absence of a trion two-quantum peak ($2T^{-}$) is modeled by maintaining equivalence of the $|T^{-}\rangle \rightarrow |2T^{-}\rangle$ and $|g\rangle \rightarrow |T^{-}\rangle$ transitions.  The unequal strength of the $LXT^{-}$ and $HXT^{-}$ peaks in the two-quantum spectrum in Fig. \ref{fig2}(c) originates from EID of the exciton transition in the presence of the trion, which is modeled by increasing the dephasing rate of the $|T^{-}\rangle \rightarrow |XT^{-}\rangle$ transition compared to the $|g\rangle \rightarrow |X\rangle$ transition.  Through EID, the $HXT^{-}$ peak destructively interferes with the $2X$ peak at the exciton + trion energy.  The same EID mechanism enhances the $HP$ amplitude compared to the $LP$ in Fig. \ref{fig2}(a), and the symmetric shape of these peaks is reproduced only when $R=0$ for the quantum pathways involving both the exciton and trion, indicating that fluctuations of their respective transition energies are uncorrelated.  MBEs also enhance the exciton peak in the one-quantum spectrum in Fig. \ref{fig2}(a) relative to the trion peak, which is weaker due the absence of interactions between trions.  The simulation demonstrates that the non-radiative Raman-like coherences in the zero-quantum spectrum are concealed by many-body effects.

In summary, coherent interactions between excitons and trions in a CdTe/CdMgTe QW have been studied using optical 2DCS.  The collection of zero-, one- and two-quantum spectra provides sufficient constraints for establishing how MBEs influence the coherent optical response of excitons, trions and a 2DEG in a QW.  Excellent agreement between density matrix calculations and the experiment is obtained, from which several conclusions can be drawn.  First, cross peaks in the spectra appear from an excitation-induced energy shift of the mixed exciton-trion state.  Second, an asymmetry in the coupling peak amplitudes indicates that the presence of trions enhances the exciton dephasing rate.  Third, the shape of the cross peaks in the rephasing one-quantum spectrum indicates that fluctuations in the exciton and trion transition frequencies are uncorrelated.  Fourth, a two-quantum coherence signal at the collective exciton + trion energy reveals that the interactions are coherent in nature.  Lastly, the absence of a trion two-quantum signal reveals that trions do not coherently interact due to their spatial separation.  These observations cannot be attributed simply to mixing of the exciton and trion wave functions mediated by the 2DEG \cite{Suris2001}, which has been a useful concept for explaining renormalization of the exciton and trion energies and oscillator-strength-stealing-phenomena; instead, the results presented here necessarily require nonlinearities arising from exciton-trion interaction effects contributing to the nonlinear response.  Optical 2DCS provides a unique perspective into the many-body effects stemming from coherent exciton-trion coupling, which we anticipate will motivate continued theoretical work based on microscopic multi-particle interactions that fully capture the many-body effects between neutral and charged particles residing in a plasma.

The authors thank R. A. Suris for helpful discussions.  The research at JILA was primarily supported by the Chemical Sciences, Geosciences, and Energy Biosciences Division, Office of Basic Energy Science, Office of Science, U.S. Department of Energy under Award $\#$ DEFG02-02ER15346, and the National Science Foundation under Award $\#$ 1125844.  The research in Germany was supported by the Deutsche Forschungsgemeinschaft.  The research in Poland was partially supported by the National Centre of Science (Poland) Award $\#$ DEC-2012/06/A/ST3/00247. S.T.C. acknowledges support from the Alexander von Humboldt Foundation.

%

\end{document}